\documentclass[12pt]{iopart}

\usepackage{graphicx}

\begin{document}

\title{Filtration of the gravitational frequency shift in the radio links
communication with Earth's satellite.}

\author{Gusev A.V.,Rudenko V.N }

\address{Sternberg Astronomical Institute, Lomonosov Moscow State University, Moscow,Russia}
\ead{valentin.rudenko@gmail.com}
\vspace{10pt}
\begin{indented}
\item[]December 2017
\end{indented}

\begin{abstract}
At present the Radioastron (RA) Earth's satellite having very elliptic orbit is used for probing of the gravitational red shift effect \cite{1,2}. Objective of this test consists in the enhancing accuracy of measurement to check the correspondence of value of the effect to Einsten's theory at one order of value better then in was done in the GP-A experiment \cite{3}. There are two H-masers in disposal, one at the board of satellite and other at the Land Tracking Station (LTS). One can compare its mutual time rate using communication radio links between RA and LTS. In contrast with the GP-A experiment there is a possibility of measurement repetition and accumulation of data in the process of RA orbital circulation. In principle it might be resulted in the increasing of the integral accuracy. In this paper we investigate the achievable accuracy in the frame of particular method of the red shift extraction associated with the techical specific of RA mission. 

\end{abstract}

% Uncomment for PACS numbers
%
% Uncomment for keywords
%\vspace{2pc}
%\noindent{\it Keywords}: XXXXXX, YYYYYYYY, ZZZZZZZZZ
%
% Uncomment for Submitted to journal title message
%\submitto{\JPA}
%
% Uncomment if a separate title page is required
%\maketitle
% 
% For two-column output uncomment the next line and choose [10pt] rather than [12pt] in the \documentclass declaration
%\ioptwocol
%

\section{Introduction}
In the papers \cite{1}, \cite{2} the problem of measurement of the gravitational redshift effect using the Radioastron satellite was considered. It was supposed that the advantage of multi-measurement opportunity arising due to the cyclical orbital motion of the apparatus provides the improvement of accuracy proportional to the square root of number of measurements on convenient orbits. Thus having H-maser standards on the board and LTS with the same quality as in the GP-A experiment \cite{3} one could reach accuracy in one order of value better then it was provided in the GP-A mission i.e. on the level of $10^{-5}$ after 100 orbit data accumulation.

Testing the gravitational redshift effect now is in the focus of active experimental research. The ongoing experiment with Galileo 5 and 6 satellites \cite{4} aims at an accuracy of $4\cdot 10^{-5}$. This experiment benefits from fortunate launch of the two navigation satellites into elliptic orbits, their stable onboard hydrogen maser clocks and the large number of ground receivers, while the principal difficulty is the problem of taking into account systematic effects. The more advanced specialized ACES mission \cite{5}, \cite{6}, with expected launch to ISS in 2018, has the goal of reaching $2\cdot 10^{-6}$.

Early several proposals for satellite missions using onboard clocks were considered in \cite{7}, \cite{8}, \cite{9}. The reason for the increased attention to such tests is due to the fact that the gravitational redshift effect is one of the cornerstones of general relativity. Its validity supports the fundamental principle of the position invariance of physical laws. A deep analysis of the current status of various experimental tests of general relativity contains in the review paper of C.M.Will \cite{10}. Details on relativistic effects in the Global Positioning System (GPS), in particular on timing and redshifts one can find in \cite{11}.

The gravitational redshift effects also are attracted for testing different theories of relativistic gravity confronting them with GR \cite{12}, \cite{13}. For such comparison the high accuracy in measuring the effect is especially important. Recently the new step toward a solid theoretical basis for the redshift tests has been made in \cite{14}, wherein the authors derive an exact expression for the general relativistic redshift in the Schwarzschild space-time and, in particular, consider the important case of elliptic orbits around the massive gravitating body.

At practice for experiments with the ``clock on satellite'' the key point is an accuracy with which the board standard frequency can be reconstructed through the signal received on the LTS. It is the question of efficiency of the filtering procedure using for cleaning the RA signal from contaminative coherent hindrances such as Doppler shift, atmospheric distortions etc.

Experience of the deep space apparatus tracking elaborated several methods or adaptive estimate-compensation algorithms of signal processing \cite{15}. However practically used algorithms mostly have an empirical character. It would be desirable to estimate the upper limit of resolution for the board standard frequency reconstruction on the base of general receipts of theory of optimal filtration for to use it as a reference ``bench mark'' forecasting the achievable accuracy in the redshift effect measurement with RA satellite.

\section{Parameters of RA mission and signal structure.}

The physical and technical charasteristic of the Radioastron mission were paresented in several papers. The detailed description one can find in \cite{16}. Here it is worth to remind that it has the very eccentric elliptical orbit around the Earth, evolving from cicle to cicle due to the gravitational influence of the Moon (and other factors), within a broad range of the orbital parameter space: the perigee altitude $(1000\,- \, 80000) km $, and apogee altitude $(270000\,- \,350000) km $. The orbital period is variated in the interval $(8\,-\,10)$ days. The gravitational frequency shift between the clock at the Earth surface and infinity has the order $\Delta f/f \approx 6.6\cdot 10^{-10}$, or $6.6\,Hz$ in absolute value at radio frequency $10\,GHz$. For the board clock the shift is modulated along the orbit: the modulation amplidue achievs $\sim 2.5 \,Hz$ at the orbits with low perigee $(1 - 10)\cdot 10^{3}km$ and is in ten time less at the majority of average orbits with perigee $\sim 50 \cdot 10^{3}km$.

The two identical hydrogen frequency standards are installed on the RA board and LTS having the minimum of Allen deviation $2\cdot 10^{-15}$ under the averaging time $\sim 1000\,s$ (production of the ``Vremya Che'' company \cite{16}). Two main carrier frequencies are used in the communication line of LTS with space apparatus: 8.4 GHz and 15 GHz; the first one so called ``pure tone'' is utilized for technical control and tuning procedures, the second serves for a trasmittion of astrophysical data registered by the space radio telescope. But in principle the both are suitable for the gravitational redshift measurement. These measurements require special observational session which mostly are incompatible with radio astronomical observations.In particular because very often its are associated with the transmitting (at $15\, GHz $) a special broadband signal with comb-like spectrum where each subtone can be considered as a separate communication channel \cite{2}.

Quality of the orbital parameters measuring is carachterized by the following values: coordinate accuracy $200\,m $ using the radio-control but $\sim 2\,cm $ through the laser ranging. The accuracy of velocity measurements was on the level $\sim 2 mm/s$.

\section{Specifics of signal receivering and processing}
A time limited quasi harmonic signal from spacecraft is received on the Land Tracking Station , heterodyned on intervening frequency and digitized. Then the quadrature components are calculated for to form observable variables : phase and phase derivative reflecting the carrier frequency of the signal packet (sample).

Some routine procedure of frequency mesurement is carried out roughly on-line at the LTS, but then a more precise estimate can be extracted off-line with the signal recorded at the finite observation time interval (example of such procedure one can find in the paper \cite{4}).

\begin{figure}[h]
%\center{
\includegraphics[width=1\linewidth]{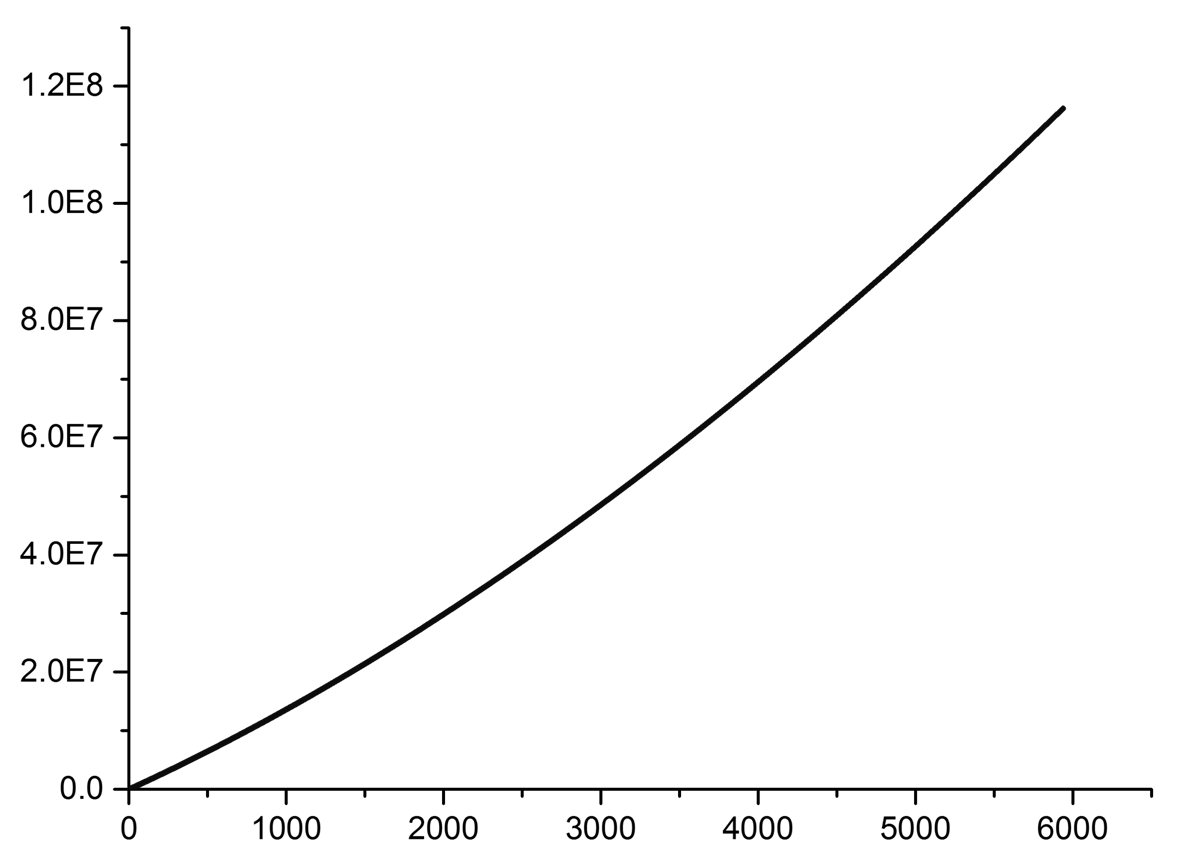}
%}
\caption{LTS. Phase after 1st order Doppler shift compensation. It is seen the nonlinear character of phase evolution for enough long observation. Y-axis -- Phase, rad; X-axis -- Session time, s.}
\label{fig:1}
\end{figure}

\begin{figure}[h]
%\center{
\includegraphics[width=1\linewidth]{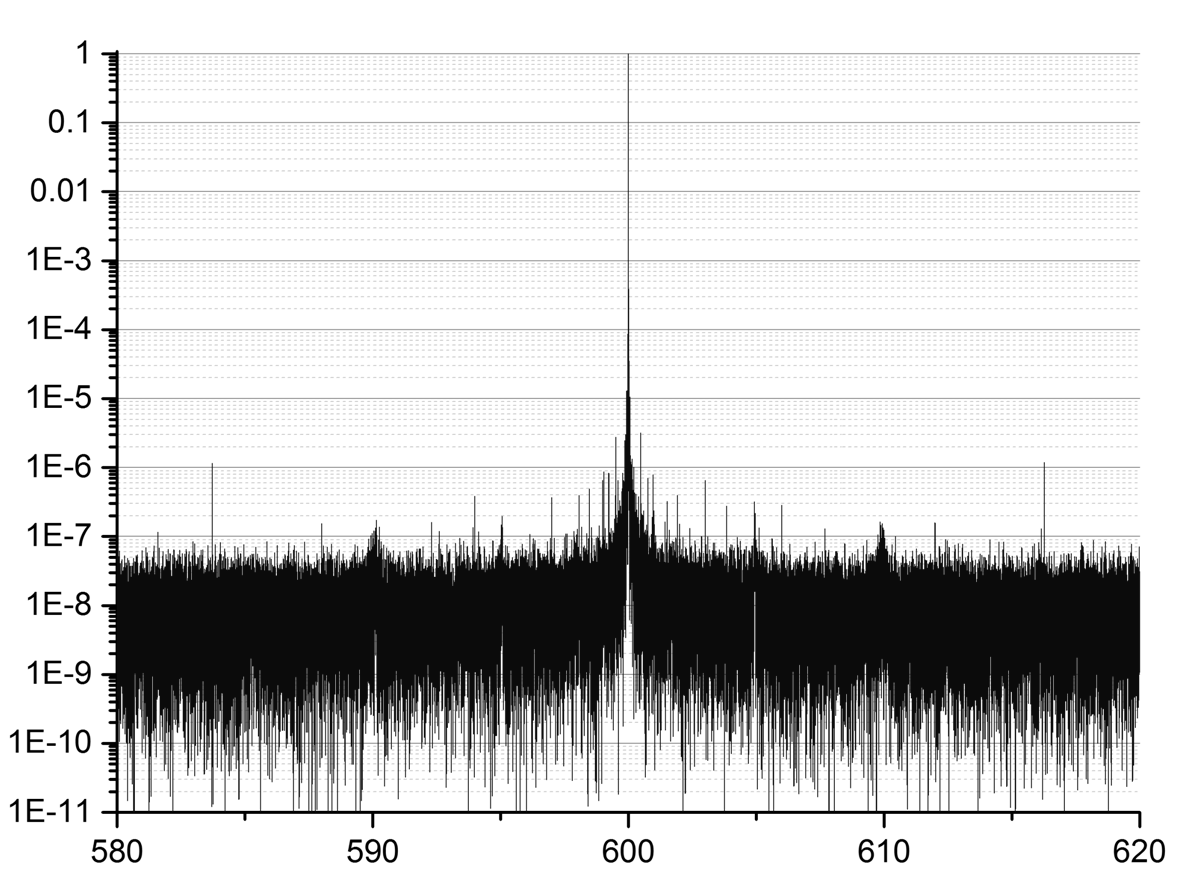}
%}
\caption{Spectra of the received signal after the ``stopped phase'' algorithm; the width of line estimate $\sim 3\cdot 10^{-4}$~Hz. Y-axis -- Normalized power (log scale); X-axis -- Frequency, Hz.}
\label{fig:2}
\end{figure}

[A typical empirical algorithm contains a number of transformations back and forth in frequency and time domains.Usually at the beginning one produces the spectrum of the signal, then the region of maximum frequency component is filtered by some spectral window and shifted to low frequency side. After that one comes back to the time domain reproducing the phase time evolution at the observational interval. Slow phase drift is estimated by LMS resulting in the regression curve and after subtraction it from the full phase one gets residual phase data called as ``stopped phase''. Derivative of the stopped phase and its spectrum provide the estimate of frequency value and the ``width of line'', i.e. the error of the frequency estimation, figures \ref{fig:1}, \ref{fig:2}.]

The two main operation regimes are foreseen for RA communication line with LTS. The first called as ``H-maser'' or ``one-way'' mode, is used for receiving RA signals on the carrier phase-locked with the board H-standard. The second called as ``Coherent'', or ``two-way'' mode is applied for signals initially sent by LTS and retrasmitted back by RA on the carrier phase-locked with the land H-standard. Such configuration provides the unique opportunity for filterring the gravitational effect from predominant Doppler shift (4 orders larger). In fact the 1st-order Doppler shift of the two-way link is twice that of the one-way downlink, but the gravitational frequency shift is absent in two-way signal. A proper combination of these signals by a radio engineering scheme can essentially eliminate the 1st-order Doppler completely retain the gravitational contribution. Partly it is true also for the troposphere and some other hindrances (initially such method of the gravitational shift filterring was realized in GP-A \cite{3}).

To clarify this point let's define preliminary the observable variables (phase and phase derivative (frequency)) formed from the input RA-signal $x(t)$ (written in the formalism of complex numbers) by the routine processing.
$$
x(t)=\mathrm{Re}\left[\widetilde{x}(t)\exp\left\{j\bar{\omega}t\right\}\right]
\eqno(1a)
$$
here the signal complex envelope read as
$$
\widetilde{x}(t)=A(t)\exp\left\{j\varphi(t)\right\}
\eqno(1b)
\label{1b}
$$
the notation $A(t)$, $\varphi(t)$ are the input signal amplitude and phase with $\bar{\omega}$ - as intervening frequency.

After digitizing with sampling time $\Delta t$ one has the digital form of the signal $x_{d}(t)$
$$
x_{d}(t)=x(t)\sum\limits_{k}\delta(t-k\Delta t)=\sum\limits_{k}x_{k}\delta(t-k\Delta t),
$$
with $x_{k}=x(k\Delta t)$ and $\Delta t\le\pi/\omega_{m}$ where $\omega_{m}$ - a maximal frequency in the spectrum $X(\omega)$ of the signal $x(t)$ i.e.

$$
x(t)\leftrightarrow X(\omega)=\Delta t\sum\limits_{k}x_{k}\exp\left\{-j\omega k\Delta t\right\},\;
|\omega|\le\omega_{m}.
$$
One has to have in the mind that $\widetilde{x}(0)=2\widetilde{X}(\bar{\omega}\pm \omega)$ with $\omega\ll \bar{\omega}$ i.e. spectrum amplitude of the complex envelope $\widetilde{x}(0)$ is twice larger the spectrum of narrow band process $\widetilde{X}(\omega)$.

It is supposed below that spectrum $\widetilde{X}(\omega)$ is concentrated in the bandwidth $(-\Omega,\Omega)$. Then according to the sampling theorem the quadrature components of the signal can be presented by the formulae
$$
\left\{
\begin{array}{cc}
A_{c}(t)=\mathrm{Re}\widetilde{x}(t)=A(t)\cos\varphi(t)=
\frac{\Omega\Delta t}{\pi}\sum\limits_{k}x_{k}\cos(k\bar{\omega}\Delta t)\mathrm{sinc}\,(\Omega{t}-k\Delta t),
\\
A_{s}(t)=\mathrm{Im}\widetilde{x}(t)=A(t)\sin\varphi(t)=
-\frac{\Omega\Delta t}{\pi}\sum\limits_{k}x_{k}\sin(k\bar{\omega}\Delta t)\mathrm{sinc}\,(\Omega{t}-k\Delta t).
\end{array}\right.
\eqno(2a)
$$
Having the quadrature components in disposal one can get the observable variables (phase and phase derivative (or frequency)) as
$$
\varphi(t)=\arctan\frac{A_{s}(t)}{A_{c}(t)}+\pi k,\,\, f(t)= \frac{d}{dt}\,\, {\varphi}(t).
\eqno(2b)
$$
Phase increment at a some interval of observation $(t_{m},t_{m+1})$ is defined by the expression:
$$
\Delta\varphi_{m}(t)=\varphi(t)-\varphi(t_{m})=
\int\limits_{t_{m}}^{t}\dot{\varphi}(t)dt,\;t_{m}\le t\le t_{m+1}=t_{m}+T_{m}.
\eqno (3)
$$
The initial phase $\varphi(t_{m})$ in our analysis below will be considered as unknown value.

Now let's come back to the algorithm of 1-st order Dopler shift compensation using data of the two operation regimes mentioned above.

Phenomemologicaly one can present a total phase increment for each separate mode as the following equations
$$
\left\{
\begin{array}{cc}
\Delta\varphi_{1w}=\Delta\varphi_{g}+ \Delta\varphi_{c}+ \Delta\varphi_{n},
\\
\Delta\varphi_{2w}=2(\Delta\varphi_{c}+ \Delta\varphi_{n}).
\end{array}\right.
$$
Here $\Delta\varphi_{g}$ - the gravitational increment; $\Delta\varphi_{c}$ - contributions of the 1st order Doppler and other coherent hindrances (slow perturbations variated on the scale of orbital motion); $\Delta\varphi_{n}$ - the stochastic phase variations produced by other type of noises. Thus the gravitational term appears only in the 1w-signal (initiated by the board H-standard). The 2w-signal (synchronized by the LTS-standard) contains only information about hindrances and noise background twicely increased in respect of the 1w communication.

The evident receipt of the gravitational effect on-line filtering consists in the substraction of data measured in two channels (both modes) simultaneously. More precisely one has to take as observable variable the combination $ \Delta\varphi_{mix}$ characterized by the following formula
$$
\Delta\varphi_{mix} = \Delta\varphi_{1w} - (1/2)\Delta\varphi_{2w} =
\Delta\varphi_{g} + \Delta\varphi_{fl}.
\eqno(4)
\label{eq:4}
$$
where the term $\Delta\varphi_{fl}$ describes the residual environmental and instrumental fluctuation background. It was metioned above this algorithm was successfully used in the experiment GP-A \cite{3}.

\begin{figure}[h]
%\center{
\includegraphics[width=1\linewidth]{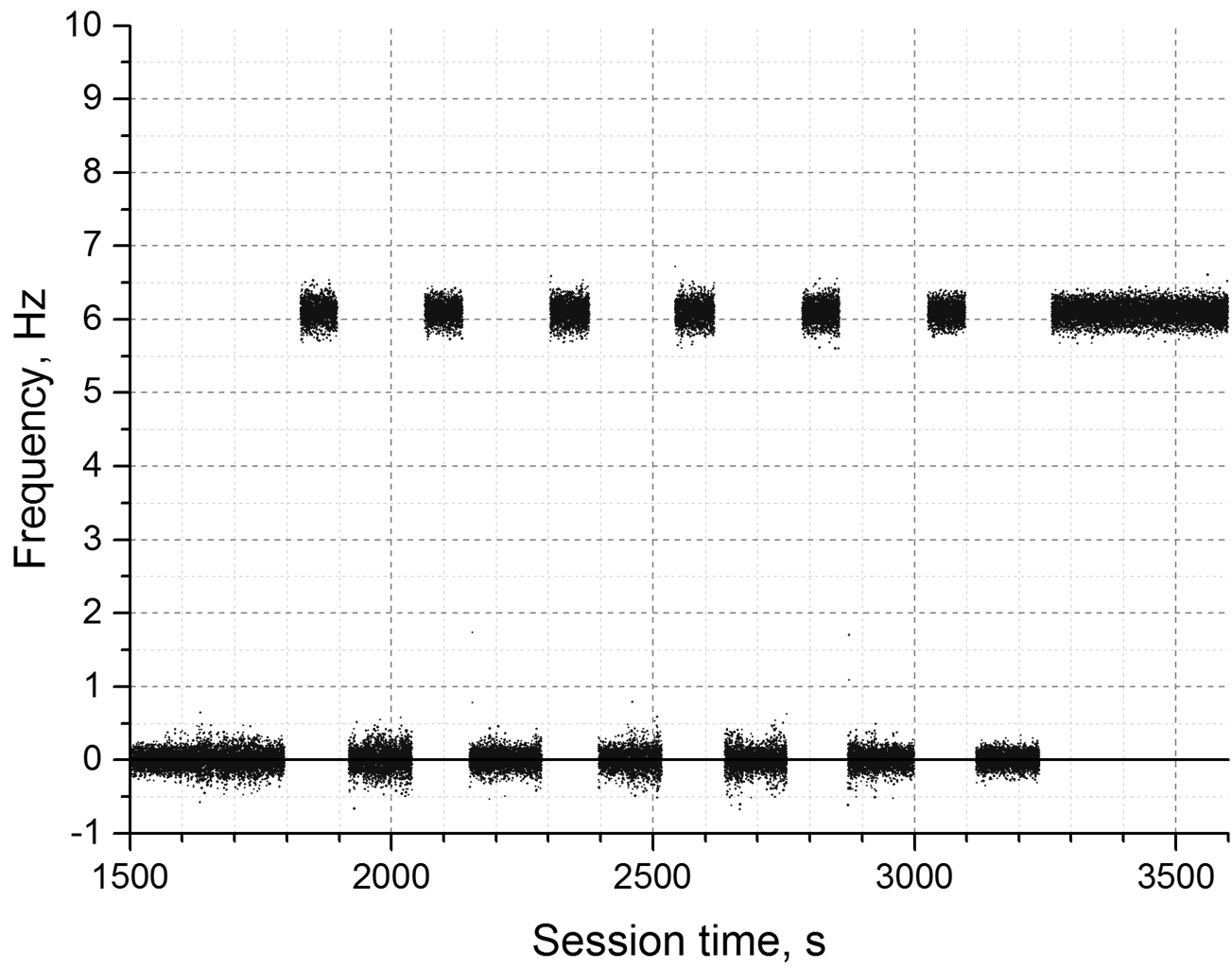}
%}
\caption{LTS Measured frequencies in the mixed mode regime after coherent hindrance compensation; the rough estimate of the redshit magnitude $\sim$6.2~Hz, distance 250~000~km.}
\label{fig:3}
\end{figure}

However this method can not be applied directly to the gravitational measurement with RA mission. The problem consists in the absence of techical ability to operate simultaneously with two communicalion modes. Then there is the only opportunity of modes alternation reswitching periodically one-way and two-way mode operation during the observational session. Example of such manner of operation at one of the RA orbit is presented at the (Fig \ref{fig:3}).

It leads to the ne�essity to perform an interpolation of experimental data of each mode at the ``empty places'' - i.e. intermediate time intervals between subsiquent measurements. Using the interpolated data one can implement the ``subtraction algorithm'' for two different modes in the same time moment. The only need is to take into account the error of interpolation inherent in such operation.

Interpolation might be performed by LMS method if the fluctuation background would be describe as the gaussian ``white noise''. This is relatively probable for the ``phase variable''. However for the frequency as the phase derivative it is not obligatory valid and one would has to apply the Likelihood algorithm to get the continuous approximation of frequency evolution.

The objective of this paper consists in the study of differential algorithms for the mixed mode regime , the estimation of its accuracy and some illustration of its efficiency at the partial experimental data of RA mission.

\section{Interpolation in the time piecewise record}
Let us take the phase ${\varphi}(t)$ as obsevable variable, then a probable model of its time behavior at the interval of observation $T$ might looks as
$$
\varphi(t)=s(t)+n(t),\;0\le t\le T,
\eqno(5a)
$$
here $s(t)$ - a slow variation of the phase associated with the satellite orbital motion and drifts of other nature; it might be approximated by polinomial time function on the order of $M$
$$
s(t)\approx s_(t,\mathbf{a})=\sum\limits_{i=1}^{M}a_{i}t^{i},
\eqno(5b)
$$
where
$\mathbf{a}=\|a_{1}\ldots a_{M}\|^{\mathrm{T}}$ are unknown parmeters and $n(t)$ - additive random background type of the gaussian white noise with the correlation function
$$
k_{n}(\tau)=\left<n(t)n(t+\tau)\right>=N\delta(\tau),\;N=\frac{1}{2Q\Delta f},
\eqno(6)
$$
where the notaion $Q$ is a signal-noise ratio in the reciption bandwidth $\Delta f$.

The order of the approximation polinomium $M$ is not given and might be choosed using a prior information about the physics of coherent hindrances or through some adaptive empirical procedure which will be describe below.

At practice the stochastic variable $\varphi(t)$ is measured at separate time intervals $(t_{m},t_{m+1})$, $m=\overline{0,L}$. The corresponded phase variations are defined by the formula (3). In general the initial value of the phase at each interval
$$
\varphi(t_{m})=\varphi_{m}=s(t_{m},\mathbf{a})+n(t_{m})
$$
has to be considered as unknown.

Now the problem of interpolation phase data at the empty record intervals can be formulated as LMS task. One has to construct the regression line for experimental ``phase peaces data'' in the polinomial form written above (5b) (i.e. one has to define the parameters $\mathbf{a}$) taking into account also optimal choice of unknown initial phases $\varphi(t_{m})$. Then the multiparametric LMS extremum equation read as
$$
\sum\limits_{m}\int\limits_{t_{m}}^{t_{m+1}}[\Delta\varphi_{m}(t)
-s(t,\widehat{\mathbf{a}})+\widehat{\varphi}_{m}]^{2}dt=\min,
\eqno(7)
$$
where $\widehat{\mathbf{a}}=\|\widehat{a}_{1}\ldots \widehat{a}_{M}\|^{\mathrm{T}}$ - the estimates of parametes $\mathbf{a}$ and $\widehat{\varphi}_{m}$ - estimates of initial phases.

At first from the extremum equation of initial phases
$$
\frac{\partial}{\partial\widehat{\varphi}_{m}}
\int\limits_{t_{m}}^{t_{m+1}}[\Delta\varphi_{m}(t)
-s(t,\widehat{\mathbf{a}})+\widehat{\varphi}_{m}]^{2}dt=0
$$

one can find its optimal estimates
$$
\widehat{\varphi}_{m}=\frac{1}{T_{m}}\int\limits_{t_{m}}^{t_{m+1}}
[\Delta\varphi_{m}(t)- s(t,\widehat{\mathbf{a}})]dt,\;T_{m}=t_{m+1}-t_{m}.
$$
At the next step solving the system of linear equations
$$
\frac{\partial}{\partial\widehat{a}_{j}}
\int\limits_{t_{m}}^{t_{m+1}}[\Delta\varphi_{m}(t)
-s(t,\widehat{\mathbf{a}})+\widehat{\varphi}_{m}]^{2}dt=0,\;j=\overline{1,M}
$$
one gets estimates of the $\mathbf{a}$ parameters from relations
$$
\sum\limits_{i=1}^{M}\widehat{a}_{i}\sum\limits_{m}T_{m}\left(\vartheta_{m}^{i+j}-
\vartheta_{m}^{i}\vartheta_{m}^{j}\right)=
\sum\limits_{m}\int\limits_{t_{m}}^{t_{m+1}}\Delta\varphi_{m}(t)\left(t^{j}-\vartheta_{m}^{j}\right)dt,\;
j=\overline{1,M},
\eqno(8)
\label{eq:8}
$$
where the notation of $\vartheta_{m}^{j}$ were introduced
$$
\vartheta_{m}^{j}=\frac{1}{T_{m}}\int\limits_{t_{m}}^{t_{m+1}}t^{j}dt=
\frac{1}{T_{m}(j+1)}\left(t_{m+1}^{j+1}-t_{m}^{j+1}\right).
\eqno(9)
\label{eq:9}
$$
Presentation of this result can be reduced to a brief vector form after the following definition
$$
\mathbf{I}\equiv[I_{ij}=\sum\limits_{m}T_{m}\left(\vartheta_{m}^{i+j}-
\vartheta_{m}^{i}\vartheta_{m}^{j}\right)]
\eqno(10)
\label{eq:10}
$$
%$$
%\left\{
%\begin{array}{cc}
%\mathbf{I}\equiv[I_{ij}=\sum\limits_{m}T_{m}\left(\vartheta_{m}^{i+j}-
%\vartheta_{m}^{i}\vartheta_{m}^{j}\right)] \\
%\mathbf{f}\equiv        [f_{j}]=\sum\limits_{m}\int\limits_{t_{m}}^{t_{m+1}}\Delta\varphi_{m}(t)
%\left(t^{j}-\vartheta_{m}^{j}\right)dt.
%\end{array}\right.
%$$
The estimates of the components of the vector $\mathbf{a}$ one can find from the equations (8) using the definitions (9), (10). In the matrix form it has the compact expression
$$
\widehat{\mathbf{a}}=\mathbf{a}+\mathbf{I}^{-1}\mathbf{f},
\eqno (11a)
$$
here the perturbative vector $\mathbf{f}$ is discribed by the right parts of the equations (8). To define it in a more clear form it is convenient to rewrite the phase evolution, given by (5a), (5b), separating the interpolation uncertainty from fluctuation error i.e.
$$
\varphi(t)= \sum\limits_{i=1}^{M}a_{i}t^{i},+ R(t)+n(t)
$$

Above we replace the ``signal'' variable $s(t,\mathbf{a})$ by its polinomial approximation of the $M$ order with the residuals $R(t)$ (systematic or interpolation error).

Then the perturbation vector $\mathbf{f}$ is devided as $\mathbf{f} = \mathbf{f_{R}} + \mathbf{f_{n}}$ where the first term is the systematic interpolation uncertainty, while the second presents the stocastic error. More in details the new verctors are
$$
\mathbf{f}_{R}=\|f_{R1}\,\ldots\,f_{Rm}\,\ldots\,\|^{\mathrm{T}},\;
f_{Ri}=\sum\limits_{m}\int\limits_{t_{m}}^{t_{m+1}}R(t)(t^{i}-\vartheta_{m}^{i})dt,
\eqno (12a)
$$
$$
\mathbf{f}_{n}=\|f_{n1}\,\ldots\,f_{nm}\,\ldots\,\|^{\mathrm{T}},\;
f_{ni}=\sum\limits_{m}\int\limits_{t_{m}}^{t_{m+1}}n(t)(t^{i}-\vartheta_{m}^{i})dt.
\eqno (12b)
$$

Thus the accuracy of estimation of the $\mathbf{a}$ parameters is defined by the interpolation and stochastic parts
$$
\delta\mathbf{a}=\widehat{\mathbf{a}}-\mathbf{a}=\mathbf{I}^{-1}(\mathbf{f}_{R}+\mathbf{f}_{n}).
\eqno (11b)
$$
The first term in this expression presents the systematic error, the second - fluctuation contribution:
$$
\delta\mathbf{a}=\delta\mathbf{a}_{n}+\delta\mathbf{a}_{R}.
$$
In the white noise
$$
\left<\mathbf{f}_{n}\mathbf{f}_{n}^{\mathrm{T}}\right>=N\mathbf{I}
$$
so that,
$$
\left<\delta\mathbf{a}_{n}\delta\mathbf{a}_{n}^{T}\right>
=N\mathbf{I}^{-1}.
\eqno (13a)
$$

The systematic or interpolation error is read as
$$
\delta a_{Rk}=\sum\limits_{i=1}^{M}I_{ik}^{-1}f_{Ri}.
\eqno (13b)
$$
or more in details
$$
|\delta a_{Rk}|\le \frac{\max\limits_{0\le t\le T}s^{(M+1)}(t)}{(M+1)!}
\sum\limits_{i=1}^{M}\sum_{m}
|I_{ik}^{-1}T_{m}(\theta_{m}^{M+i+1}-\vartheta_{m}^{M+1}\vartheta_{m}^{i})|.
$$
here the $s^{(M+1)}(t)$ is $(M+1)$ derivarive of the signal part.

\section{Gravitational shift extraction in the mixed mode regime }
Now let's consider the key question of RA gravitational data processing algorithm. With which accuracy the gravitational frequency shift can be estimated (measured) using the mixed mode regime? 

The principle receipt of ``Red Shift'' extraction with on-line compensation of coherent hindrances was formulated and realized in GP-A experiment: it's simultaneous receivering of the ``one way'' and ``two way signals'' and measuring its difference at the output of corresponded hardware radiotechnical circuit of the land tracking station \cite{3}. Specifics of the RA radio tracking system allows to do similar procedure only through the off-line manner using switchings between two mentioned modes. The interpolation procedure to reconstruct continious phase evolution in each fixed mode was described in the section above. Here we study the problem of gravitational shift extraction in the mixed mode regime.

Let us note phase time variations in each mode as $ \varphi_{1w}(t)$ for the H-maser mode (one way) and $\varphi_{2w}(t)$ for the coherent mode (two way). The both have the signal and noise components
$$
\left\{
\begin{array}{cc}
\varphi_{1w}(t)=s_{1w}(t)+n_{1}(t), \\
\varphi_{2w}(t)=s_{2w}(t)+n_{2}(t),
\end{array}\right.\;0\le t\le T,
$$
Precise polinomial interpolation of the ``signal variable'' $s(t)$ teoretically requires an infinit rank number. Cut off at the polinomial rank $M$ is accompanied by the residual term $R(t)$ which reflects the interpolation (or systematic) error
$$
\left\{
\begin{array}{cc}
s_{1w}(t)=\sum\limits_{k=1}^{\infty}a_{k}t^{k}=\sum\limits_{k=1}^{M}a_{k}t^{k}+R_{1}(t), \\
s_{2w}(t)=\sum\limits_{k=1}^{\infty}b_{k}t^{k}=\sum\limits_{k=1}^{M}b_{k}t^{k}+R_{2}(t). 
\end{array}\right.
\eqno (15)
 $$
Then follow the on-line compensation algorithm of \cite{3}, \cite{17} one can write for frequencies at the output of differential link 
 $$
 \frac{d}{dt}\left[s_{1}(t)-\frac{1}{2}s_{2}(t)\right]=f_{g}(t)+f_{e}(t)+0(v/c)^{4},
\eqno (16 )
 $$
 here 
 $$
 f_{g}(t)=\sum\limits_{k=0}^{\infty}\alpha_{k}t^{k}
 $$
 - the gravitational shift frequency,
 $$
 f_{e}(t)=\sum\limits_{k=0}^{\infty}\beta_{k}t^{k}
 $$
- the shift produced by coherent hindrances; its can be estimated using measured data of the orbit parameters and modeled atmospheric characteristics \cite{1}, \cite{2}, \cite{17}.

Above the auxillary parameters $\alpha_{k}$, $\beta_{k}$ were introduced for the sake of formulae simplicity.

Its are coupled with the main coefficients
$$
\alpha_{k}+\beta_{k}=(k+1)\left(a_{k+1}-\frac{1}{2}b_{k+1}\right).
\eqno (17)
$$
Through the LMS processing (see section\,4) these coefficients gets their estimations $\widehat{a}_{k}(M)$,\,$\widehat{b}_{k}(M)$ associated with the choosed polinomium rank $M$ and corresponded errors $\delta a_{k}$,\,$\delta b_{k}$ i.e.
$$
\left\{
\begin{array}{cc}
\widehat{a}_{k}(M)=a_{k}+\delta a_{k}, \\
\widehat{b}_{k}(M)=b_{k}+\delta b_{k},
\end{array}\right.\;k=\overline{1,M},
$$
Having the principle interest in the measurement of the gravitational frequency shift one can get from (17) the following expression for the $a_{k}$ coefficients 
$$
\alpha_{k}+\delta\alpha_{k}=(k+1)\left[\widehat{a}_{k+1}(M)-\frac{1}{2}\widehat{b}_{k+1}(M)\right]-
\beta_{k},\;k=\overline{0,M-1},
\eqno (18a) 
$$
with
$$
\delta\alpha=(k+1)\left[\delta a_{k+1}(M)-\frac{1}{2}\delta b_{k}(M)\right].
\eqno (18b)
$$
the parameters $\beta_{k} $ in (18) are considered here as known values (controlled by auxillary measurements of coherent hindrances).

Thus one can conclude that the error of the redshift estimate depends on the rank $M$ of polinomium approximation. The total error contains contributions of stochastic (noise) uncertainty ( 12b),(13) and systematic (interpolation) error ( 12a),(14). Increasing the rank power surpress the systematic error, but enhances the stochastic (noise) contribution ( a larger number of noisy term one has to take into account). The optimal solution of this contradiction is similar to the general recommendation from the theory of ``ill posed problems'' \cite{18}. One has to take the polinomial rank $M\ge M^{*}$ from the condition: the systematic (interpolation) error is equal (or less) the stochastic one. The empirical adaptive algorithm of the $M^{*}$ search for is presented into Appendix 1. If this condition is fulfiled then
$$
\begin{array}{cc}
\delta a_{k}(M)=\delta a_{n,k}(M)+\delta a_{R.k}(M)\simeq\delta a_{n,k}(M), \\
\delta b_{k}(M)=\delta b_{n,k}(M)+\delta b_{R.k}(M)\simeq\delta b_{n,k}(M). 
\end{array}
$$
At last the final formula for the total error of the gravitational frequency shift in polinomium approximation is defined by the Fisher matrix $\bf {I}$ \,for the polinom coefficients (10),(13), i.e. 
$$
M\ge M^{*}:\;\delta\alpha_{k}\simeq\delta\alpha_{n,k},
$$
$$
\left<\delta\alpha_{i}\delta\alpha_{k}\right>\simeq N(i+1)(k+1)
\left\{\left[I_{i+1k+1}^{-1}(M)\right]_{1w}+\frac{1}{4}\left[I_{i+1j+1}^{-1}(M)\right]_{2w}\right\}.
\eqno(19)
$$

\section {Numerical estimate on the example of three intervening }
As some illustration of the general algorithm extracting of the redshift value from experimental data of RA in the mixed mode regime let's present below the numerical result of calculation of achivable accuracy for the simple case of 3 reswitching between the modes.

1) First of all one needs to know the spectrum density of the receiver noise in the model of ``white noise in the limited reciption bandwidth''. The technical characteristics of the LTS in ``Puschino'' are the following: the equivalent noise temperature of the receivering tract, including antenna paraboloid, less then 100 K into bandwidth $\sim 1 KHz $. At the last measuring amplifier the signal/noise ration for typical communication session was $Q \sim 10^{6} $ in the measuring band width $ 25 Hz$. It results in the spectral density (6) $N\approx 0.5\cdot 10^{-8}\,Hz^{-1}$.

2) Duration of the separate time interval $T_{m}=T_{1}=\mathrm{const} \simeq 600 s$

3) moments of rewitching $t_{m}=2(m-1)T_{1}$

4) number of reswitching $m=\overline{1,3}$.

5) adaptive calculation of the approximation polinom results in $M=7$.

6) elements of the Fisher matrix $\mathbf{I}$ were calculated according to the formulae (9), (10) 
$$
[I_{ik}(M)]=T_{1}\sum\limits_{m}(\vartheta_{m}^{i+k}-\vartheta_{m}^{i}\vartheta_{m}^{k})
$$
$$
\vartheta_{m}^{i}=\frac{T_{1}^{i}}{(i+1)}[(2m-1)^{i+1}-(2m-2)^{i+1}].
$$
The final estimate of the frequency shift measurement error (uncertainty of the polnomial coefficient $\alpha_{0}$) results in 
$$
\sqrt{\left<\delta\alpha_{0}^{2}\right>}\simeq2,3\cdot10^{-5}Hz.
$$
It means that the relative accuracy of the redshift extraction through the two mode adaptive algorithm with the RA data might reach the value $2.3\cdot 10^{-5}Hz/ 6\,Hz \approx 4\cdot 10^{-6}$. It does not contradict with the goal to improve the GP-A result \cite{3} inspite of the absence of on-line compensation filterring.

\section{Conclusions}
Above we have studied a possible optimization of the filtering procedure of the gravitational redshift effect extracting from the communication lines between RA satellite and LTS. The on-line filtering processing used in the GR-A experiment is not applicable due to the technical reason: an absence of simultaneous using the both foreseen one-way and two-way operational modes. One has to deal only with the regime of their intervening. It means that a direct subtraction of the coherent hindrances in the communication channel is replaced by some off-line estimate-compensation algorithm; (such situation is typical for many spacecraft missions: the board generator used for a scientific information transmitting, but the land one used for driving commands and as a duplicate generator (with two-way mode) in the case of destruction of the board one; generally the joint operation of the both generators is not foreseen). In this situation the question of possible loose of sensitivity (or resolution degradation) becomes very important one and needs in a special investigation. In this paper we have considered the variant of gravitational frequency shift measurement through data of the mixing mode regime solving the interpolation error problem in competition with the noise error. Specific results of our analysis might be briefly formulated as follow:

- System of linear equations was found (8) for estimation of the interpolation polinomium coefficients in the mixing mode regime. Regression curve is calculated on the base of all available pieces of data with optimization along the unknown initial phase of each piece. After that a choice of polinom coefficients was also optimized.

- Stochastic error of interpolation coefficients was calculated in the frame of the additive Gaussian white noise model.

- Some adaptive algorithm was elaborated (independent on physical nature of noises) for estimation of the minimal rank of approximation polinomium with the condition that the interpolation error does not exceed the fluctuation one.

- Preliminary tests of the method was carried out with particular data received during of the RA gravitational session in the mixed mode regime. Results does not contradict on the goal of achieving the accuracy of redshift measurement one order better then in GP-A experiment after $\sim 20$ session data accumulation

\section*{Acknowledgement}
Authors would like gratitude members of radiotechnical group of the RA team: Birukov A.V., Kovalenko A.V., Smirnov A.I. for the help in understanding of the signal-noise parameters of RA apparatus and Litvinov D.A. for many fruitful discussions. This work was supported by the national grant RSCF - 17-12-01488

\section*{Appendix 1}
\subsection{Empirical adaptive algorithm of the optimal polinomium rank $M$ determination}
Objective of the adaptive algorithm is to find the rank $M^{*}$ of approximation polinomium under which the stochastic error $\delta\mathbf{a}_{n}(M)$ exceeds the interpolation one $\delta\mathbf{a}_{R}(M)$. On mathematical language this condition can be written as
$$
\left|\Delta\widehat{s}(t|M)\right|\le
\sqrt{\left<\Delta s_{n}^{2}(t|M)\right>},
\eqno (A1)
$$
where in the left side there is the difference of the signal term estimates with ranks $M$ and $M+1$, i.e.
$$
\Delta\widehat{s}(t|M)=\widehat{s}(t|M+1)-\widehat{s}(t|M).
$$
The relation between the estimate $\widehat{s}(t|M)$ and uncertainty of interpolation coefficients $\delta{a}_{k}(M)$ looks like
$$
\widehat{s}(t|M)=\sum\limits_{k=1}^{M}\widehat{a}_{k}(M)t^{k}=
s(t|M)+\sum\limits_{k=1}^{M}\delta a_{k}(M)t^{k},
$$
In its turn the right part of (A1) contains the stochastic variance 
$\left<\Delta s_{n}^{2}(t|M)\right>$ of the signal term estimates with ranks $M$ and $M+1$. It is convenient to introduce the stochastic variable $\xi_{n}(t|M)$,
$$
\xi_{n}(t|M)=\sum\limits_{k=1}^{M}\delta a_{n,k}(M)t^{k}.
$$
Then one can clarify the fluctuation variance in the right part of (A1)
$$
\Delta s_{n}(t|M)=\xi_{n}(t|M)-\xi_{n}(t|M+1),
$$
$$
\left<\Delta s_{n}^{2}(t|M)\right>=\left<\xi_{n}^{2}(t|M)\right>+\left<\xi_{n}^{2}(t|M+1)\right>-
2\left<\xi_{n}(t|M)\xi_{n}(t|M+1)\right>
$$
To avoid the complex formulae with tripple sign of summarizing it is enough to present here the expressions for variance of $\xi_{n}(t|M)$ variable and correlation matrix elements of stochastic errors $\delta a_{n,k}(M)$, i.e.
$$
\left\{
\begin{array}{cc}
\left<\xi_{n}^{2}(t|M)\right>=N\sum\limits_{i=1}^{M}\sum\limits_{k=1}^{M}
I_{ik}^{-1}(M)t^{i+k}, \\
\left<\delta a_{n,i}(t|M)\delta a_{n,j}(t|M+1)\right>=
N\sum\limits_{k=1}^{M}\sum\limits_{m=1}^{M+1}I_{ki}^{-1}(M)I_{mj}^{-1}(M+1)I_{ij},
\end{array}\right.
$$
Thus the empirical adaptive algorithm consists from the follwing steps: for the initial (choosed) rank $M=2,3,\ldots$ one calculates the threshold levels $\left<\Delta s_{n}^{2}(t|M)\right>$; then its has to be compared with values $\left<\Delta s_{n}^{2}(t|M)\right>$ composed from the experimental (measuring) data. Under condition 
$$
|\Delta\widehat{s}(t|M=M^{*})|\le \sqrt{\left<\delta s_{n}^{2}(t|M=M^{*}\right>}
$$
the decision $M\simeq M^{*}$ has to be accepted.

\end{document}